\documentclass[pre,showpacs,preprint]{revtex4}
\usepackage{graphicx}
\usepackage{epsfig}

\begin{document}

\title{Thermodynamics of  ultracold trapped gases. Generalized mechanical variables, equation of state and heat capacity.}

\author{Nadia Sandoval-Figueroa and V\'{\i}ctor Romero-Roch\'{\i}n\footnote{Corresponding author}}

\email{nadia@fisica.unam.mx, romero@fisica.unam.mx}

\affiliation{Instituto de F\'{\i}sica, Universidad Nacional
Aut\'onoma de M\'exico. \\
Apartado Postal 20-364, 01000 M\'exico,
D.F. Mexico.}

\begin{abstract}

 The thermodynamics framework of an interacting quantum gas trapped by an arbitrary external potential is reviewed. We show that for each confining potential, in the thermodynamic limit, there emerge ``generalized" volume and pressure variables ${\cal V}$ and ${\cal P}$, that replace the usual volume and hydrostatic pressure of a uniform system. This scheme is validated with the derivation of the virial expansion of the grand potential. We show that this approach yields experimentally amenable procedures to find the equation of state of the fluid, ${\cal P} = {\cal P}({\cal V}/N,T)$ with $N$ the number of atoms, as well as its heat capacity at constant generalized volume $C_{\cal V} = C_{\cal V}({\cal V},N,T)$. With these two functions, all the thermodynamics properties of the system may be found. As specific examples we study weakly interacting Bose gases trapped by harmonic and by linear quadrupolar potentials within the Hartree-Fock approximation. Comparisons with experimental results of a $^{23}$Na ultracold gas are also presented. We claim that this route should provide an additional and useful tool to analyze both the thermodynamic variables of a trapped gas as well as its elementary excitations.

 \end{abstract}


\date{\today}
\maketitle

\section{Introduction.}

The realization of quantum ultracold trapped gases\cite{Anderson,Davies,Bradley,Greiner,Bartenstein,Thomas1,Magalhaes,Ziewerlein,Hulet,Thomas2,Shin} has opened new test grounds of  both known and novel states of matter. This, in turn, has vigorously stimulated first principles analysis of the physics of many body systems\cite{Butts,Dalfovo,Perali,RR1,RR2,RR3,RR4,Yukalov1,Yukalov2,Yukalov3,DeSilva,Gubbels,Chien,Ho,Drummond,Thomas3}. Among the many fundamental aspects of these fluids, a ubiquitous characteristic stands out, namely, the fact that these systems are spatially inhomogeneous due to the confining non-uniform external potential of the trap. Hence, usual theoretical tools, concepts, and the concomitant physical intuition that has been developed for uniform systems confined by rigid-wall vessels, must be adapted or completely reformulated to deal with the ensuing nonuniformities and the presence of the trap itself. The main purpose of the present article, in the light of this fact, is to review the emergence of ``new" and proper mechanical variables for confined fluids that modify the costumary way  one is use to deal with their thermodynamic  properties.

The main differences on which we base our discussion, between the thermodynamics of non-uniform trapped gases and that of uniform systems, is that while the hydrostatic pressure $p$ of the fluid becomes a local quantity, the volume $V$ that the gas occupies is no longer precisely defined. That is, these two variables are not anymore the appropriate thermodynamic variables to describe the mechanical equilibrium of the inhomogenous trapped fluid. The role of the usual $p$ and $V$, however, is replaced by  a unique pair of variables for each confining potential, called {\it generalized} pressure ${\cal P}$ and volume ${\cal V}$. With the aid of these two variables, thermodynamics can then be fully used and exploited as the usual tool to characterize and describe the state of the gas. It is our contention that useful and typical properties of fluids, such as equations of state, heat capacities and compressibilities, have not been introduced in the description of the physics of the trapped gases so far, because of the lack of the appropriate thermodynamic variables. It is our goal to help filling this gap. As we shall discuss and show here, these properties can be extracted from the knowledge of the density profile and other simple properties of the trapped gases. We emphasize that thermodynamics, besides characterizing a macroscopic system,  also yields information regarding both its microscopic interactions and/or its elementary excitations.

The present article builds in previous works and results dealing with a quantum Bose gas trapped in a harmonic potential\cite{RR1,RR2,RR3,RR4} and in comparisons with experiments in  ultracold Sodium gases\cite{Silva,Henn,Henn2}. Here, we extend the analysis to any arbitrary external confining potential and show specific results for Bose gases trapped in both  harmonic and linear quadrupolar traps.

We shall proceed in the following way. First, we describe the general system under study - an interacting fluid confined by an external, inhomogeneous potential $V_{ext}(\vec r)$ - and we provide several specific examples to introduce the generalized volume ${\cal V}$ and  pressure ${\cal P}$. We discuss their physical interpretation and the corresponding thermodynamic limit in which these variables emerge. We briefly mention additional thermodynamic variables that may arise from the external potential itself and from the possibility of varying the interatomic interactions externally; the latter is now a reality being widely exploited\cite{Ziewerlein,Hulet,Thomas2,Shin}. We show how the equation of state, ${\cal P} = {\cal P}({\cal V}/N,T)$, and the heat capacity at constant volume ${\cal V}$, $C_{\cal V} = C_{\cal V}({\cal V},N,T)$, can be measured within the current experimental setups.
We then explictly deal with a confined quantum gas including interatomic interactions, and we derive the virial expansion of the grand potential to validite our approach. As a corollary of our treatment we show that the so called ``local density approximation" (LDA) follows within the appropriate thermodynamic limit of the confining potentials
\cite{Garrod,MarchioroI,MarchioroII}. Because the density inhomogeneities appear at macroscopic length scales, it may be expected that LDA should apply, nevertheless we do not make any explicit assumption on the density profiles and it is thus reassuring to find that the virial expansion when applied to confined quantum fluids yields LDA. As we shall discuss, however, care must be taken when using it. We shall see that while it gives a procedure to calculate thermodynamic properties of a confined inhomogeneous fluid, it does not imply that the local states of the trapped fluid are thermodynamic states of the corresponding homogenous fluid.

We then devout a section to apply our general framework to a confined weakly interacting Bose gas. We study a gas confined both by a harmonic and a linear quadrupolar traps within the Hartree-Fock approximation. We calculate the phase diagram ${\cal P}-T$ and the heat capacity at constant volume. A brief discussion on the nature of the normal gas to superfluid transition is presented. We show then a  comparison of these predictions with experimental results of Bagnato et al.  on Sodium gases\cite{Henn}.

\section{Thermodynamic variables of trapped gases}

The system consists of $N$ identical
atoms or particles of mass $m$ with Hamiltonian
\begin{equation}
H_N = \sum_{i=1}^N \frac{\vec p_i^2}{2 m} + \sum_{i<j} u(|r_{ij}|) +
\sum_{i=1}^N V_{ext}(\vec r_i) .\label{HN}
\end{equation}
We assume additive pairwise potentials
but the analysis may be extended to arbitrary interatomic
interactions. Below, we shall consider the second quantized version of the above Hamiltonian to explicitly include the scattering length. The external potential of the trap $V_{ext}(\vec r)$ confines
the system. To serve this purpose, it should have at least one minimum
 and must obey that $V_{ext}(\vec r) \to \infty$
for $|\vec r| \to \infty$. For rigid-wall containers it is
costumary not to write down the potential. Here, we include
it as $V_{ext}(\vec r) = 0$ if
$\vec r$ is within the volume $V$ enclosed by the rigid walls and
$V_{ext}(\vec r) = \infty$ if $\vec r$ is
outside of it. Typical examples of traps of atomic gases are
$V_{ext}(\vec r) = (1/2) m (\vec \omega \cdot \vec r)^2$ a harmonic
potential, such as in Ref.\cite{Anderson}, and
$V_{ext}(\vec r) = |\vec A \cdot \vec r|$
 a linear quadrupolar potential\cite{Silva};
but one can
consider any confining potential such as a P\"oschl-Teller\cite{PT,Mirena}
$V_{ext}(\vec r) = V_0/\cos(\vec \gamma \cdot \vec r)$. This last case serves as an example of a potential that while it gives rise to a generalized volume, it also introduces an additional intensive variable, namely $V_0$.
We write these potentials to exemplify the appropriate
thermodynamic variables as well as the thermodynamic limit for each case.

To illustrate how the generalized variables emerge and how the thermodynamic limit is to be taken, we shall deal here first with a classical ideal gas. Below, we shall treat an interacting fluid and we shall verify the correctness of the identification of the variables given here. Consider, therefore, a system given by the Hamiltonian (\ref{HN}) with no interatomic interactions, i.e. $u(|r_{ij}|) \equiv 0$. Assume the system is in thermodynamic equilibrium at temperature $T$. The canonical partition function is,
\begin{equation}
Z(T,N,{\cal V},\eta) =  \frac{1}{h^{3N} N!} \int d^{3N} p \> \int d^{3N} r \> \exp\left[{-\beta \sum_{i=1}^N \left(\frac{\vec p_i^2}{2 m} +  V_{ext}(\vec r_i)\right) }\right] ,
\end{equation}
 where $\beta = 1/kT$.  We assume the external confining potential to be of the form $V_{ext} = V_{ext}(x/l_x,y/l_y,z/l_z,\eta)$ where the quantities $l_i$ do not necessarily have units of length and $\eta$ stand for other parameters, such as $V_0$ in the P\"oschl-Teller potential above. Integration of the partition function yields
\begin{equation}
Z(T,N,{\cal V},\eta) = \frac{1}{N! \lambda_T^{3N}} \left(\zeta(\beta, \eta) {\cal V}\right)^N ,
\end{equation}
where $\lambda_T = h/(2\pi mkT)^{1/2}$ is de Broglie thermal wavelength, ${\cal V} = l_x l_y l_z$ is the generalized volume and the function $\zeta(\beta,\eta)$ is defined by
\begin{equation}
\zeta(\beta, \eta) {\cal V} = \int \> e^{-\beta V_{ext}(\vec r)} \> d^3 r. \label{zeta}
\end{equation}
Helmholtz free energy is found with $F = - kT \ln Z$ and, after taking the limit $N \to \infty$, yields
 \begin{equation}
 F(N,T,{\cal V},\eta) = -NkT \left( \ln \left[\frac{{\cal V} \zeta(\beta,\eta)}{N \lambda_T^3}\right] + 1\right).
 \end{equation}
 For the free energy per particle, $F/N$, to remain finite in the thermodynamic limit, $N \to \infty$, it must be required that the ``generalized" volume diverges, i.e. ${\cal V} \to \infty$, keeping constant the ``density"  $N/{\cal V}$. As it will fully justified below,  ${\cal V}$ is an {\it extensive} thermodynamic variable. The generalized volume certainly is proportional to the actual {\it average} volume that the system occupies, $\bar V \sim \zeta(\beta,\eta) {\cal V}$. This average volume, however, is not an independent thermodynamic variable since it depends on the temperature. Moreover, it is not a correct variable since the actual volume that the system occupies is, in general, unbounded. Nevertheless,  the thermodynamic limit ${\cal V} \to \infty$ indeed implies that the volume of the system becomes arbitrarily large.

For the particular external potentials here considered one finds, ${\cal V} = V$ for rigid walls,
${\cal V} = 1/\omega_x \omega_y \omega_z$ for the harmonic potential, ${\cal V} = 1/A_x A_y A_z$ for the quadrupolar potential, and
${\cal V} = 1/\gamma_x \gamma_y \gamma_z$ for the P\"oschl-Teller potential. Likewise, we find  $\zeta(\beta) = 1$ for rigid walls,
$\zeta(\beta) = (2\pi /\beta m)^{3/2}$ for the harmonic potential,
$\zeta(\beta) = 8\pi /\beta^3$ for the quadrupolar potential,
and
$\zeta(\beta,V_0) = 4\pi \int_{-\pi/2}^{\pi /2} x^2 \exp [- \beta V_0 / \cos x] dx$
for the
P\"oschl-Teller potential. For the harmonic case, it has been known for quite a long time that the thermodynamic limit is the one here presented\cite{Groot}. As we shall see, while the role of the generalized volume is completely analogous to that of the usual volume in homogeneous systems, the thermodynamic properties of the different confined fluids show very strong variations on their temperature dependences due to the function $\zeta(\beta,\eta)$.

As it will be fully justified below with the virial expansion for an {\it interacting} gas confined in an arbitrary potential, the generalized volume is a bona-fide extensive variable. Therefore, there exists an {\it intensive} variable, conjugate to the volume ${\cal V}$, that we call the generalized pressure ${\cal P}$ and given by
\begin{equation}
{\cal P} = - \left(\frac{\partial F}{\partial {\cal V}}\right)_{N,T} . \label{forP}
\end{equation}
Here, $F = F(N,T,{\cal V})$ is Helmoltz free energy including interatomic interactions. By a simple calculation one obtains,
\begin{eqnarray}
{\cal P}
&=& \frac{1}{3 \cal V} \left< \sum_{i=1}^N \vec r_i \cdot \nabla_i V_{ext}(\vec r_i) \right> \nonumber \\
&=&  \frac{1}{3 \cal V} \int \> \rho(\vec r) \> \vec r \cdot \nabla  V_{ext}(\vec r) \> d^3 r ,\label{PV}
 \end{eqnarray}
where in the first line the average is performed in the corresponding ensemble and in the second line we have introduced the density profile $\rho(\vec r)$. The last equality is a very useful tool, as we shall insist throughout this paper: It gives a direct way to calculate the generalized pressure with the sole knowledge of the external potential $V_{ext}(\vec r)$, the density profile $\rho(\vec r)$ and the temperature $T$. That is, the equation of state ${\cal P} = {\cal P}({\cal V}/N,T)$ is a measurable quantity.
In Section V we shall use this result for the calculation of the phase diagram of a confined interacting Bose gas.

It is important to point out that the identification of the generalized pressure is not only a formal one but it has a clear physical meaning. From the thermodynamic definition of ${\cal P}$, Eqs.(\ref{forP}) and (\ref{PV}), one sees that the product ${\cal PV}$ equals (1/3) of (minus) the virial of the external force. Hence, one can recall that mechanical equilibrium in a fluid is given by Pascal law,
\begin{equation}
\nabla \cdot \tilde P(\vec r) = - \rho(\vec r) \nabla V_{ext}(\vec r) ,\label{Pascal}
\end{equation}
where $\tilde P(\vec r)$ is the pressure tensor of the fluid. One expects the
pressure tensor to be a local quantity, $\tilde P(\vec r) = p(\vec r) \tilde 1$,
where $\tilde 1$ is the unit tensor and $p(\vec r)$ the local hydrostatic
pressure, barring phase-separated states within the confined fluid. By calculating the virial of the right hand side of Eq.(\ref{Pascal}), and after integrating by parts, one finds
\begin{eqnarray}
\int {\rm Tr} \tilde P \> d^3 r &=&  \int d^3 r \>
\rho(\vec r) \> \vec r \cdot \nabla V_{ext}(\vec r) \nonumber \\
&=& 3 {\cal PV} . \label{tr}
\end{eqnarray}
That is, ${\cal P}$ for a non-uniform fluid confined by a given external potential plays the same thermodynamic role as the hydrostatic pressure $p$ in a uniform fluid: it is the quantity that bears the information that the fluid is in mechanical equilibrium. It is somewhat puzzling to realize that expression (\ref{PV}) yields only an identity for the rigid-wall case and does not give a calculational tool for the hydrostatic pressure. The latter needs the knowledge, at least for pairwise interatomic interactions, of the two-body density correlation function\cite{RW}. Here, we find that for inhomogeneous systems knowledge of the one-body density suffices.

We note that with the above identification the change in free energy is
\begin{equation}
dF = - SdT - {\cal P} d{\cal V} + \mu dN \label{df}
\end{equation}
with $S$ the entropy and $\mu$ the chemical potential, provided the rest of the intensive parameters of the external potential, as well as those of the interatomic potential $u(r_{ij})$, see below, remain constant. For instance, if the external potential has additional intensive parameters $\eta$, such as $V_0$ in the P\"oschl-Teller potential, and these are also externally modified, there are additionally changes  in the free energy given by\cite{Mirena}
\begin{equation}
 \Gamma d\eta ,
 \end{equation}
 where $\Gamma$ stands for  conjugate  {\it extensive} variables to $\eta$,
 \begin{equation}
 \Gamma =  \left(\frac{\partial F}{\partial \eta}\right)_{T,N,{\cal V}} .
\end{equation}
  The variables $\eta$ enter through the function $\zeta(\beta, \eta)$ only, see Eq.(\ref{zeta}). From the above relationship one finds that an external potential can give rise to as many thermodynamic variables (and their conjugates) as the number of parameters needed to specify it. However, there is only one variable that plays the role of the volume. For simplicity, we shall assume that the additional intensive variables $\eta$ of the external potential remain constant.

 The measurement of the heat capacity $C_{\cal V}$ should also be achievable within the current experimental setups. The proposal of this measurement consists of an adiabatic compression or expansion. We note first that the ultracold trapped gases are actually isolated and confined by magnetic or optical traps\cite{Anderson,Davies,Bradley,Greiner,Bartenstein,Thomas1,Magalhaes,Ziewerlein,Hulet,Thomas2,Shin}. Therefore, a slow change of the confining potential, namely, of the generalized volume ${\cal V}$, should give rise to an increase or decrease of the temperature $T$ depending on whether the fluid is compressed or expanded. Since the generalized volume and temperature are measurable in the current experiments, the quantity $(\partial {\cal V}/\partial T)_{N,S}$ can, therefore, be calculated. The corresponding heat capacity can then be evaluated using then the following thermodynamic relationship,
\begin{equation}
C_{\cal V} = - T \left(\frac{ \partial {\cal P}}{\partial T}\right)_{{\cal V},N} \left(\frac{\partial {\cal V}}{\partial T}\right)_{N,S} .
\label{cv}
\end{equation}
We note that previous knowledge of the equation of state is needed for the calculation of the second factor on the right hand side of (\ref{cv}). However, the measurements of the equation of state and of the quantity $(\partial {\cal V}/\partial T)_{N,S}$ correspond to two different sets of experiments.

It is also of interest here to mention that one of the most interesting and novel aspects of the current ultracold gases is the fact that the interatomic interaction potential may be modified externally by means of magnetic fields
\cite{Ziewerlein,Hulet,Thomas2,Shin}. This, in turn, traduces into an external modulation of the scattering length. That is, the scattering length becomes a (intensive) thermodynamic variable itself and, therefore, there appears a thermodynamic extensive variable conjugated to it. To be explicit, we write down the Hamiltonian in a second quantized version and consider a contact interatomic potential,
 \begin{equation}
 H = \sum_{n} \epsilon_n a_n^\dagger a_n + U {\sum_{jklm}}^\prime a_j^\dagger a_k^\dagger a_l a_m
 \end{equation}
 where $n$ and $\epsilon_n$ stand for the eigenstates and eigenvalues of the three dimensional one-particle Hamiltonian in the presence of the external potential. $a_n^\dagger$ and $a$ are creation and annihilation operators. The ``prime" in the second sum refers to the restrictions introduced by assuming an isotropic two-body potential. The coupling parameter is $U = 4\pi \hbar^2 a/m$, with $m$ the atom mass and $a$ the scattering length\cite{Dalfovo}.  It has already been recognized, however, that the relevant quantity is $1/U$ rather than $U$\cite{Tan1,Tan2}. Therefore, since $U$ is clearly an intensive quantity, there exists an extensive variable ${\cal C}$ given by,
 \begin{eqnarray}
 {\cal C} &=& - \left(\frac{\partial F}{\partial 1/U}\right)_{T,N,{\cal V},\eta} \nonumber \\
 &=& \left< U^2 {\sum_{jklm}}^\prime a_j^\dagger a_k^\dagger a_l a_m \right> .
 \end{eqnarray}
 This quantity has been called the {\it contact}\cite{Tan2,Bratten} and it is also related to the large momentum tail of the particles momentum distribution.  Thus, this yields an additional term in the change of free energy given by
 \begin{equation}
 - {\cal C} d \left(\frac{1}{U}\right) .
 \end{equation}
In the current literature ${\cal C}$ has been introduced as the {\it adiabatic}
 change of the internal energy $E$ with respect to $1/U$\cite{Tan2}. From the previous analysis, one finds that such a definition is actually incorrect. Since ${\cal C}$ is the isothermal change of the free energy with respect to $1/U$, standard thermodynamics\cite{LL} indicates that in the microcanonical ensemble the correct relationship is
 \begin{equation}
 \frac{1}{U} = \left(\frac{\partial E}{\partial {\cal C}}\right)_{S,N,{\cal V},\Gamma} .
 \end{equation}
 A full study of the variables ${\cal C}$ and $1/U$ is of fundamental importance, but its is out of the scope of the present article.
Nevertheless, as an instance of the relevance of the mechanical variables with these studies, we mention here the recent interest on the behavior of Fermi gases ($^{40}$K and $^{6}$Li) near the unitarity limit where the scattering length diverges, i.e. the limit $1/U \to 0$, because it appears that thermodynamics becomes universal, that is, independent of the interatomic interactions; see e.g. Refs.\cite{Thomas1,Ziewerlein,Hulet,Thomas2,Shin,Ho,Drummond,Thomas3}. In this region, the gas behaves as an ideal one in the sense that it obeys the ideal virial relationship, which for a harmonic trap reads $E = 2N <V_{ext}>$, with $E$ the internal energy. The connection to the present work is that $N<V_{ext}>$ for a harmonic trap is $3/2 {\cal PV}$, whether the gas behaves as ideal or not, see Eq.(\ref{PV}) and Ref.\cite{RR1}. That is, the quantity that has been {\it measured} in Refs.\cite{Thomas1,Thomas2,Drummond,Thomas3} is precisely the generalized pressure for the harmonic trap. And indeed, if the universality hipothesis is correct\cite{Ho}, using the virial theorem for ideal gases for arbitrary potentials, yields the following relation that should be obeyed in the unitarity region,
\begin{equation}
E = < \sum_{i=1}^N V_{ext}(\vec r_i)> + \frac{3}{2}{\cal PV} .
\end{equation}
Away from the unitarity limit, this equation is no longer valid but the measurement of ${\cal P}$, and of
the heat capacity as we described above, can be performed to obtain the thermodynamics of those states.

\section{Virial expansion for arbitrary confining potentials.}

With the identification of the generalized variables and the corresponding thermodynamic limit in hand, we now turn to the virial expansion of an interacting quantum gas. The classical case appears as the limit of high temperatures, as we shall indicate it. We extend the analysis described in advanced textbooks on statistical physics\cite{Mayer,Haar,Blatt}. With this expansion we shall validate the thermodynamic variables introduced in the previous section. Again, we assume the system is in thermodynamic equilibrium at temperature $T$ and we analyze it in the grand canonical ensemble. We thus consider a chemical
potential $\mu$ whose value may be found by imposing a given number of
particles $N$. The grand potential is given by,
\begin{equation}
\Omega = - kT \ln \sum_{N = 0}^\infty e^{\beta \mu N} \> {\rm Tr}^\prime e^{-\beta H_N} ,
\label{Omega}
\end{equation}
where
\begin{equation}
{\rm Tr}^\prime \> e^{- \beta H_N} = \frac{1}{ N!}  \int d^{3N} r  \> W_N(\vec r_1, \vec r_2, \dots , \vec r_N)
\end{equation}
and
\begin{equation}
W_N(\vec r_1, \vec r_2, \dots , \vec r_N) = \sum_{P} \epsilon^P < \vec r_1, \vec r_2, \dots , \vec r_N | e^{- \beta H_N}|\vec r_{1P}, \vec r_{2P}, \dots , \vec r_{NP}>  .
\end{equation}
The sum is over all permutations of $1,2, ... , N$ and $\epsilon = \pm 1$ for bosons or fermions.

To find the virial expansion, equation (\ref{Omega}) is first rewritten as,\cite{Mayer,Haar,Blatt}
\begin{equation}
-\beta \Omega = \sum_{n=1}^\infty \> e^{ \beta \mu n} \> \frac{1}{n!} \> I_n, \label{Ome-vir}
\end{equation}
where the functions $I_n$ are given by
\begin{equation}
I_n = \int d^{3n} r \>U_n(\vec r_1, \dots, \vec r_n),
\end{equation}
and, in turn,  the Ursell functions are found from the hierarchy, first order,
\begin{equation}
U_1(1) = W_1(1) ,
\end{equation}
second order,
\begin{equation}
U_2(1,2) = W_2(1,2) - U_1(1) U_1(2) , \label{U2}
\end{equation}
third order,
\begin{eqnarray}
U_3(1,2,3) &=&  W_3(1,2,3) - U_1(1) U_2(2,3) -U_1(2) U_2(1,3) - U_1(3) U_2(1,2)- \nonumber \\
&&U_1(1)U_1(2)U_1(3) .
\end{eqnarray}
and so on.

The virial expansion follows after finding the value of each contribution $I_n$ in the thermodynamic limit. We do this now for a general confining external potential $V_{ext}(\vec r)$. We proceed by systematically calculating $I_n$ order by order and then generalize it to $I_n$. We have done so from $I_1$ to $I_4$. Although lengthy, the corresponding calculations are straightforward and we explicitly present in the Appendix the case $I_2$ only. Next we discuss the results.

The calculation of $I_1$ is very simple but serves to indicate an important point:
\begin{equation}
I_1 =
\int d^3 r < \vec r | e^{- \beta H_1} | \vec r > ,
\end{equation}
where the one-particle Hamiltonian $H_1$ is given by
\begin{equation}
H_1 = {\vec p^2 \over 2m} + V_{ext}(\vec r) . \label{H1}
\end{equation}
In the thermodynamic limit, equivalent to the use of the so-called semiclassical density of states\cite{Bagnato87}, one readily finds
 \begin{equation}
I_1 = {1 \over \lambda_T^3} \zeta(\beta) {\cal V}  .\label{I1}
 \end{equation}
The same expression would be obtained if the system obeyed classical mechanics. As we indicate below, this result is always obtained for the center of mass motion of the $n$-particle cluster involved in $I_n$.

Following the explicit calculation of $I_2$ shown in the Appendix suffices to see how to find $I_n$. The key is in the separation of center of mass and relative coordinates. This change of variables is generally, $
\vec R = \frac{1}{n} (\vec r_1 + \vec r_2 + \dots + \vec r_n)$, $\vec r^{(1)} = \vec r_1 - \vec r_2$,
$\vec r^{(2)} = \vec r_2 - \vec r_3$, $\dots$, $\vec r^{(n-1)} = \vec r_{n-1} - \vec r_n$, with their canonical conjugate momenta.  The main assumption is that, in the thermodynamic limit ${\cal V} \to \infty$, the following approximation is correct,
\begin{equation}
V_{ext}(\vec r_1) + V_{ext}(\vec r_2) + \dots + V_{ext}(\vec r_n) \approx n V_{ext}(\vec R) ,\label{terlimn}
\end{equation}
where $(\vec r_1, \vec r_2, \dots, \vec r_n)$ are to be given in terms of the variables $(\vec R,  \vec r^{(1)} ,\vec r^{(2)} ,\dots, \vec r^{(n-1)} )$ by the above transformation. This approximation separates the center of mass motion from the relative ones of the corresponding $n$-particle cluster. The ensuing motion of the center of mass is always quasiclassical and its contribution to $I_n$ is proportional to $\zeta(n\beta) {\cal V}/\lambda_T^{3n}$, see Eq.(\ref{I1}). Within the same limit, the contribution from the relative coordinates yields the usual quantum virial coefficients $b_n(T)$, the same as those calculated in the uniform case, i.e. they are universal for all confining potentials. Thus, one generally finds
\begin{equation}
I_n = \frac{\cal V}{\lambda_T^{3n}} \zeta(n\beta) \>b_n(T).\label{In}
\end{equation}
The validity of the above procedure requires, first of all, that the interatomic interactions must be ``short-range", namely, decaying faster than $1/r^3$, otherwise the virial coefficients $b_n$ do not exist\cite{Garrod,Mayer,LL}. For high temperatures, in the classical regime, one finds that the intermolecular potential must vanish for lengths $r \gg \sigma$, with $\sigma$ the range of such a potential. At low temperatures the bound is set up by either the thermal de Broglie wavelength or the scattering length $a$\cite{LLQM}. If the gas behaves as an ideal one, the relevant length is de Broglie wavelength. In any case, as long as the relative coordinates remain bounded by a {\it finite} quantity, however large, one can take the limit of very large volumes ${\cal V} \to \infty$ and separate the motion of the center of mass from the relative motions of the involved $n$ particles. It is of pedagogical interest to mention that the present approximations are always tacitly performed in the uniform case of rigid-wall potentials, by neglecting boundary terms after the corresponding change of variables.

Summarizing, we find that in the thermodynamic limit the grand potential can be written in general as,
\begin{equation}
\Omega = - kT {\cal V} \sum_{n=1}^\infty \frac{e^{\beta \mu n}}{n!} \frac{\zeta(n \beta)}{\lambda_T^{3n}} b_n(T)
\label{Ob}
\end{equation}
with $b_1 = 1$. This expression for the grand potential validates the introduction of the mechanical variables ${\cal P}$ and ${\cal V}$, as we now verify.
The number of particles $N$ and the entropy $S$ can be calculated from (minus) the
partial derivatives of $\Omega$ with respect to $\mu$ and $T$ respectively.
$\Omega$, $N$ and $S$ are found to be homogeneous first order functions
of ${\cal V}$, and  this implies that ${\cal V}$ must be an extensive variable and justifies the thermodynamic
limit as used above. Since it should be obeyed that the conjugate variable is ${\cal P} = -(\partial \Omega/ \partial {\cal V})_{T,\mu}$, it follows that $\Omega = - {\cal PV}$, as it should. Thus,  the generalized pressure is read off (\ref{Ob}).

It is interesting to note that the most important difference of the grand potential between a given arbitrary external potential and the homogeneous case is the function $\zeta({\beta})$ rather than the generalized volume ${\cal V}$. The latter enters in the same way for any potential, including the rigid-walls case; that is, it gives rise to the intensive quantities formed between the extensive variables $N$, $S$, $E$, etc. and ${\cal V}$, that remain finite in the thermodynamic limit, i.e. $N/{\cal V}$, $S/{\cal V}$, $E/{\cal V}$, etc. However, as it is now well established, the temperature dependence of the thermodynamic variables is very different and unique for each external potential. The function $\zeta({\beta})$ gives rise to those differences.

To illustrate the use of Eq.(\ref{Ob}), we apply it to an {\it ideal} quantum gas. From the analysis in the Appendix and their corresponding value for third and fourth orders, one finds that the quantum ideal virial coefficients are given by
\begin{equation}
b_n^{(0)} = \epsilon^{n+1} \> \frac{n!}{n^{5/2}} \> \lambda_T^{3(n-1)} .
\end{equation}
Hence, the grand potential for an ideal quantum gas can be written as
\begin{equation}
-\beta \Omega =  \frac{\cal V}{\lambda_T^3} \sum_{n=1}^\infty \> e^{ \beta \mu n} \> \zeta(n\beta)\>\frac{\epsilon^{n+1}}{n^{5/2}} . \label{omeiq}
\end{equation}
This formula can be directly compared with the corresponding ones for, say, the rigid walls potential ${\cal V} = V$ and $\zeta(n \beta) = 1$, or the harmonic potential ${\cal V} = 1/\omega^3$ and $\zeta(n\beta) =   (2\pi kT/nm)^{3/2}$. The ``textbook"  formulae for these potentials are,\cite{Pethick}
\begin{equation}
-\beta \Omega  = \frac{V}{\lambda_T^3} \frac{1}{\Gamma(5/2)}\int_0^\infty \frac{x^{3/2} dx}
{e^{x-\beta\mu} - \epsilon}
\end{equation}
for rigid walls, and
\begin{equation}
-\beta \Omega  = \left(\frac{kT}{\hbar \omega}\right)^3 \frac{1}{\Gamma(4)}\int_0^\infty \frac{x^{3} dx}
{e^{x-\beta\mu} - \epsilon}
\end{equation}
for a 3D isotropic harmonic potential. Expansion of the integrals of these last two equations in powers of $e^{\beta \mu}$ yield the virial expansion, Eq.(\ref{omeiq}).

For completeness of our presentation, we write down the first few terms  of the so-called virial expansion of the equation of state ${\cal P} = {\cal P} (N/{\cal V},T)$ for low densities $N/{\cal V}$. This can be done by finding $N = N(\mu,T,{\cal V})$ from Eq.(\ref{Ob}) and inverting it term by term to yield $\mu = \mu(N/{\cal V},T)$, then, substituting the result into ${\cal P} = {\cal P}(\mu,T)$:
\begin{equation}
{\cal P}(\frac{N}{\cal V},T) = \frac{N}{\cal V} kT \left[1 - \frac{1}{2} \frac{\zeta(2\beta)}{\zeta^2(\beta)} b_2(T) \frac{N}{\cal V}
+ \left(\frac{\zeta^2(2\beta)}{\zeta^4(\beta)} b_2^2(T) - \frac{2}{3} \frac{\zeta(3\beta)}{\zeta^3(\beta)}b_3(T)\right)
\left(\frac{N}{\cal V}\right)^2 + \dots \right] .
\end{equation}
Once again, we remark that the functions $\zeta(\beta)$ make all the difference. Since in some instances one can refer the calculation of the virial coefficients to a diagramatic expansion\cite{Mayer}, one finds that the diagrams sum up differently for different potentials. We also recall that this type of virial expansion was used in Ref.\cite{Silva} to fit experimental data from a gas of Sodium atoms in a quadrupolar potential.

\section{A note on the ``local density approximation".}

A corollary from the virial expansion above, Eq.(\ref{Ob}), is that the validity of the  ``local density approximation"
follows right away from
the corresponding expressions for $\Omega$, $N$ and $S$. That is, suppose a uniform gas confined by the
rigid-wall external potential, ${\cal V} = V$ and $\zeta(n \beta) = 1$.
Define the grand potential per unit  volume $\omega(\mu,T) = \Omega/V$,
the number of particles per unit volume (particle density) $\rho(\mu,T) = N/V$ and
the entropy per unit volume $s(\mu,T) = S/V$. Now consider the same system but trapped by an external potential $V_{ext}(\vec r)$. Its thermodynamic properties may then be found by implementing the ``local density approximation'': take $\omega$, $\rho$ and $s$ of the homogenous case and make those functions
per unit volume to be their  ``local" densities
$\omega(\vec r)$, $\rho(\vec r)$ and $s(\vec r)$ in the presence
of the given external potential, by replacing the
chemical potential $\mu$ by the ``local" chemical potential
$\mu_{local}(\vec r) = \mu - V_{ext}(\vec r)$. It turns out that integration of $\omega(\vec r)$,
$\rho(\vec r)$ and $s(\vec r)$ over
all space yield the {\it exact} expansions for $\Omega$, $N$
and $S$, in the presence of $V_{ext}(\vec r)$, as given by Eq.(\ref{Ob}) and its
derivatives. That is, one finds that LDA procedure gives rise to exact results. We recall that the validity of LDA for classical and quantum systems in this limit was rigorously proved in Refs.\cite{Garrod,MarchioroI} and \cite{MarchioroII}, respectively.

The above description does show that in the thermodynamic limit the system is locally homogenous and that ``locally" actually means in length scales large  compared with those of interatomic interactions. It is in this latter connection that LDA is largely used without the need of further justification. There is a warning, however, that must be raised when using LDA. It may appear that if one is able to find {\it any} thermodynamic variable $q$ for a homogenous system and express it in terms of the chemical potential $\mu$ and temperature $T$, namely $q = q(\mu,T)$, its local counterpart when in the presence of an external potential $V_{ext}(\vec r)$ is simply $q(\vec r) = q(\mu_{local}(\vec r),T)$. This, in general, is incorrect; its is strictly justified for $\Omega/V$, $N/V$ and $S/V$ only . It is incorrect, for instance, for the internal energy and other free energies, except $\Omega$, as well as for other functions such as heat capacities. This statement can be verified by using expression (\ref{Ob}) for the Grand Potential of a confined fluid.
Thus, the fact that the system is locally homogeneous does not imply that the local states of the confined fluid are thermodynamic states of the corresponding homogeneous system, i.e. $q_{local}(\vec r) \ne q(\mu_{local}(\vec r),T)$ in general.

To verify the previous statement, let us consider the internal energy density $e = E/V$ of a homogenous system. From the knowledge of $p (= - \omega)$, $s$ and $\rho$, the energy may be calculated as $e = T s - p + \mu \rho$. Thus, one can find $e = e(\mu,T)$. Then, consider a confining potential $V_{ext}(\vec r)$ and implement LDA. Presumably, the internal energy of the trapped fluid would be $E = \int e(\mu_{local}(\vec r),T) \> d^3 r$. This, however, is not the correct value of the internal energy of the fluid in the trap. The use of Eq.(\ref{Ob}) allows us to verify it by explicitly calculating $E$ for a trapped fluid. That is, an alternative form is to use the Grand Potential $\Omega$ for the fluid in the trap, as given by Eq.(\ref{Ob}), with the appropriate variables ${\cal V}$ and $\zeta(\beta)$. Then, calculate the internal energy of the trap as $E = TS -{\cal PV} + \mu N$. One finds a different result. This can be checked very simply by considering an ideal {\it classical} gas, i.e. use $b_1 =1$ and $b_n = 0$ for $n > 1$ in Eq.(\ref{Ob}). The (incorrect) result by the first route is $E = 3NkT/2$ for all traps, while the (correct) result from the second route is
\begin{equation}
E = NkT \left(\frac{3}{2} + \frac{T}{\zeta(\beta)} \frac{d\zeta(\beta)}{dT}\right) .
\end{equation}
 It is not difficult to show, using Eq.(\ref{Ob}), that the {\it local} internal energy density is actually given by
\begin{equation}
e_{local}(\vec r) = e(\mu_{local}(\vec r),T) + V_{ext}(\vec r) \rho(\mu_{local}(\vec r),T) .
\end{equation}
That is, integration of $e_{local}(\vec r)$ yields the correct internal energy of the trapped fluid, while integration of $ e(\mu_{local}(\vec r),T)$ does not. This result can be physically rationalized by arguing that $e(\mu,T)$ considers kinetic and {\it interatomic} interactions only and, therefore, the addition of the external energy density is required. For other functions the discrepancy is more difficult to assess. For instance, one can find expressions for the specific heat at constant volume using the homogeneous thermodynamics, say $C_V/V = c_V(\mu,T)$. Again, even if $e_{local}$ is used, LDA expression for $c_V(\mu_{local}(\vec r),T)$ does not integrate to the correct value found directly from the grand potential for the non-uniform fluid $\Omega(\mu,T,{\cal V})$, Eq.(\ref{Ob}). Actually, it is not even clear if a local specific heat is a meaningful quantity. That is, the heat capacity at constant volume translates locally into an specific heat at constant density; however, if a confined fluid is heated up, keeping the same number of atoms without changing the trap, its density profile changes. Hence, it is not possible to keep the local density constant while heating the system up. For the overall confined fluid, heat capacities are certainly meaningful but these must be calculated or measured either at constant ${\cal V}$ or ${\cal P}$.

\section{Equation of state and heat capacity of weakly interacting Bose gases.}

The chief importance of correctly identifying the generalized pressure and volume, resides in its use
as a tool to characterize a given system. For the simple case of a one-component gas and for a fixed interatomic interaction, there are only two independent thermodynamic variables, say, the temperature $T$ and the molar, or per particle, generalized volume $v = {\cal V}/N$. Therefore, at least two further functions of these variables should be measured or calculated independently, in order to obtain the full thermodynamics of the system. We choose the equation of state ${\cal P} = {\cal P}(v,T)$ and the specific heat at constant generalized volume $C_{\cal V}/N = c_v(v,T)$. As we have explained in Section II, these two quantities should be very easily measured in the current experiments of ultracold gases.

We now turn our attention to the calculation of the equations of state and heat capacities of a weakly interacting Bose gas confined in a harmonic and in a linear quadrupolar potential, $V_{ext}(\vec r) = (1/2) m (\vec \omega \cdot \vec r)^2$ and
$V_{ext}(\vec r) = |\vec A \cdot \vec r|$ respectively, within the Hartree-Fock approximation\cite{Goldman,Dalfovo,Pethick}. This is a self-consistent calculation for the density profiles of the thermal and the condensate densities, $\rho_{th}(\vec r)$ and $\rho_0(\vec r)$, that leads to the following set of equations\cite{Goldman},
\begin{equation}
\rho_{th}(\vec r) = \frac{1}{\lambda_T^3} g_{3/2}(\beta\left[\mu - V_{ext}(\vec r) - 2U\rho_{th}(\vec r) -
2U\rho_0(\vec r)\right]) \label{one}
\end{equation}
and
\begin{equation}
\rho_0(\vec r) = \frac{1}{U} \left(\mu - V_{ext}(\vec r) - 2U\rho_{th}(\vec r) \right) , \label{two}
\end{equation}
with the constraint that the number of particles is a given value $N$,
\begin{equation}
N = \int \> \rho_{th}(\vec r) \> d^3 r + \int \> \rho_0(\vec r) \> d^3 r . \label{three}
\end{equation}
In Eq.(\ref{one}), $g_{3/2}(\alpha)$ is the usual Bose function $g_n(\alpha)$ for $n = 3/2$. Equation (\ref{two}) is to be undertstood valid for values when the right-hand-side is positive or zero. As a matter of fact, this is how the normal to Bose-Einstein condensation (BEC) or  superfluid transition is identified, i.e. given the temperature $T$, the transition occurs for the value of the chemical potential below which the condensate density $\rho_0(\vec r)$ is different from zero. Equation (\ref{two}) is the Gross-Pitaevskii equation in the thermodynamic limit where the kinetic energy term may be safely neglected. The above set of equations suffers essentially from the fact that it does not consider the expected Bogoliubov excitations at very low temperatures\cite{Yukalov3}. However, it should be fine for temperatures near the transition\cite{Goldman}.

As it is clear from the above set equations, their solution yields the density profile $\rho(\vec r)$ and the chemical potential $\mu$ for given values of the temperature $T$ and the generalized volume $v = {\cal V}/N$. The value of generalized pressure ${\cal P}(v,T)$ is found from Eq.(\ref{PV}) and, together with $\mu(v,T)$, one can further find the molar Helmholtz potential $f = F/N$ as $f(v,T) = - {\cal P}v + \mu$. The molar entropy  $s = (\partial f/\partial T)_v$ follows and, therefore, the specific heat $c_v = T (\partial s/\partial T)_v$. Our results are summarized in Figs. 1 to 4.

\begin{figure}
  \begin{center}
    \includegraphics[scale=1]{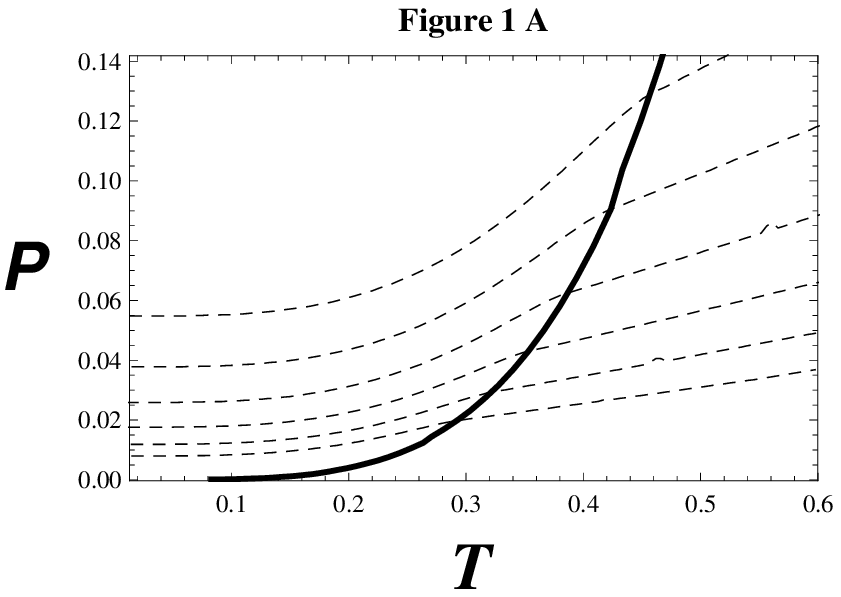}
  \end{center}
\end{figure}

\begin{figure}
  \begin{center}
    \includegraphics[scale=0.7]{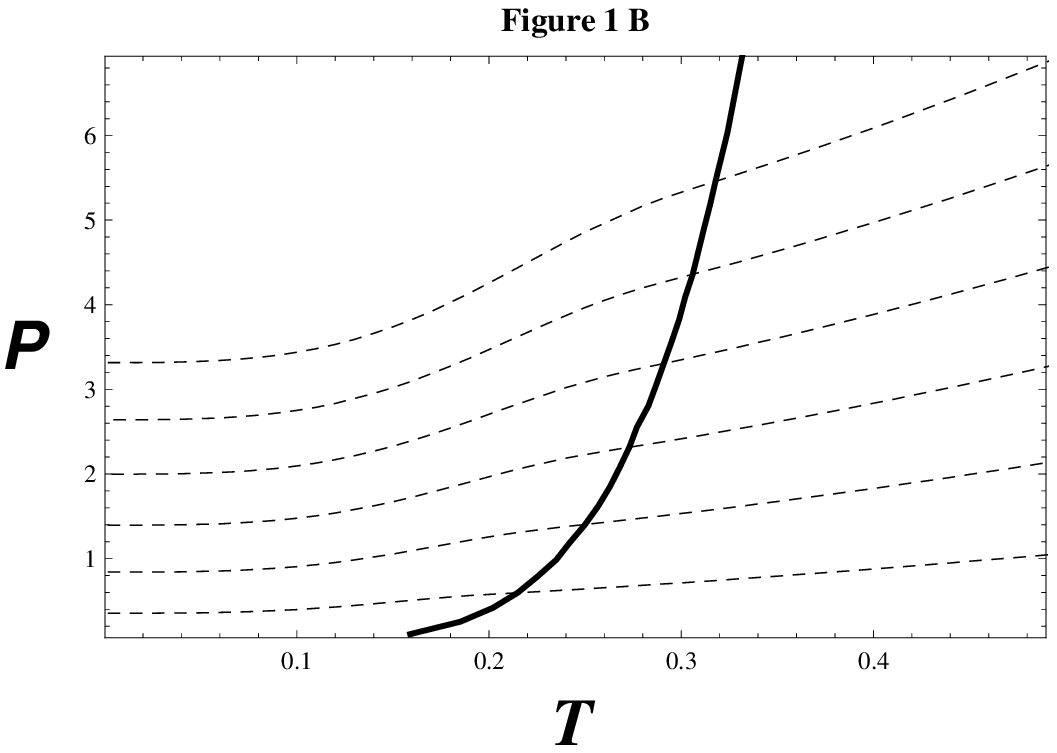}
  \end{center}
\end{figure}

\begin{figure}
  \begin{center}
    \includegraphics[scale=1]{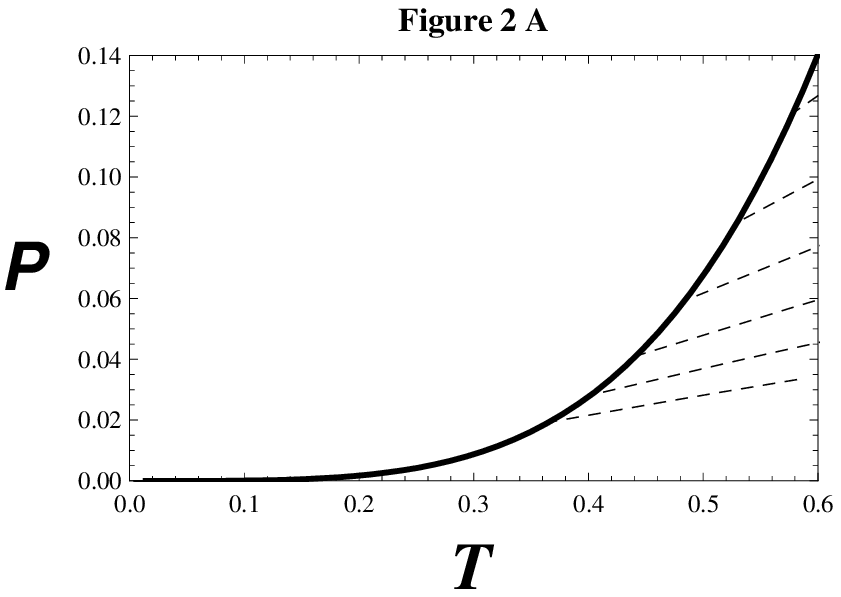}
  \end{center}
\end{figure}

\begin{figure}
  \begin{center}
  \includegraphics[scale=1]{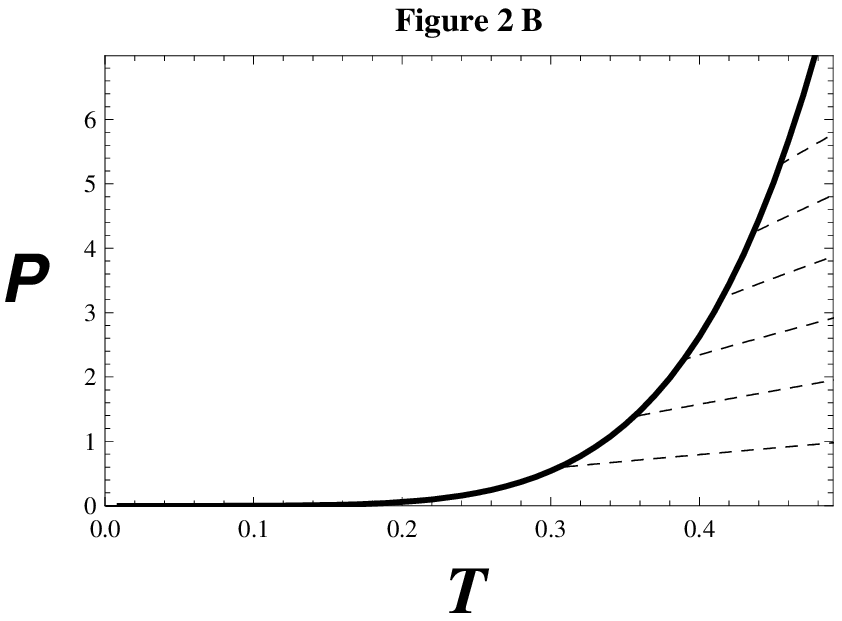}
  \end{center}
\end{figure}

Figures 1 and 2 show a few isochores ($v =$ const) of the equation of state for both external potentials, comparing the ideal case with the interacting Hartree-Fock approximation. Note that although the generalized pressures are quantitatively different, even with different units, their qualitative behavior is essentially the same. We make the following comments. First, in the ideal case the transition BEC line indicates that the pressure vanishes as $T \to 0$. That is, just as in the uniform case\cite{LL}, the condensate exerts no pressure. This is clearly changed once interactions are included: the pressure of the condensate is no longer zero, and even at $T = 0$ the interactions give rise to a remnant pressure. Second, the transition temperature, different for each isochore, is shifted down in the interacting case with respect to the ideal one. This downshift or the transition temperature is in agreement with results of more general theories of {\it trapped} Bose  gases\cite{Yukalov1}. Moreover, this is an effect due solely to the interactions and not related to finite size effects\cite{Dalfovo,Minguzzi}. And, lastly, the transition line in the interacting case, as shall be further described below, marks a smooth normal gas to superfluid transition, different to BEC where discontinuities in the second derivatives of the free energy are encountered; in the interacting case, up to second derivatives - and appears that to higher order as well - the free energy is continuous.

In Figures 3A and 3B we show the specific heat as a function of temperature, for a given isochore, for both potentials. Again the qualitative behavior is the same. Once more, we see that the transition temperature in the interacting case, marked with an arrow, is lower than that of the ideal case. But more interestingly, one finds that at the transition temperature the specific heat does not show its maximum value, but rather its {\it minimum}, and that the transition is continuous. Thus, it shows sign of being neither a first order nor a critical transition. We understand that the present is a mean-field calculation and, as mentioned above, perhaps not the best description of a superfluid; however, mean-field theories typically yield incorrect quantitative results but do not change the order of the transition. The origin of the continuity of all the thermodynamic properties may be traced back to the behavior of the condensate fraction. This is exemplified in Figure 4 where we compare the condensate fraction $N_0/N$ of the ideal with the interacting case. Below $T_c$, the ideal condensate fraction is $N_0/N = 1 - (T/Tc)^{3/2}$ for the uniform case, $N_0/N = 1 - (T/T_c)^{3}$ for the harmonic trap, and $N_0/N = 1 - (T/T_c)^{9/2}$ for the linear quadrupolar potential. Above $T_c$, $N_0/N = 0$. Thus, the transition in the ideal case has a discontinuity in the derivative. However, for the interacting case, as shown in Figure 4, there appears that this transition is completely smooth, with no discontinuity or singularity at all. Although not shown here, the corresponding isothermal compressibility of the trapped gases, $\kappa_T = - 1/{\cal V}(\partial {\cal V}/\partial {\cal P})_T$, is also continuous showing no sign of any critical fluctuations. This is also in agreement with general results concerning trapped fluids\cite{Yukalov1}.

\begin{figure}
  \begin{center}
  \includegraphics[scale=0.7]{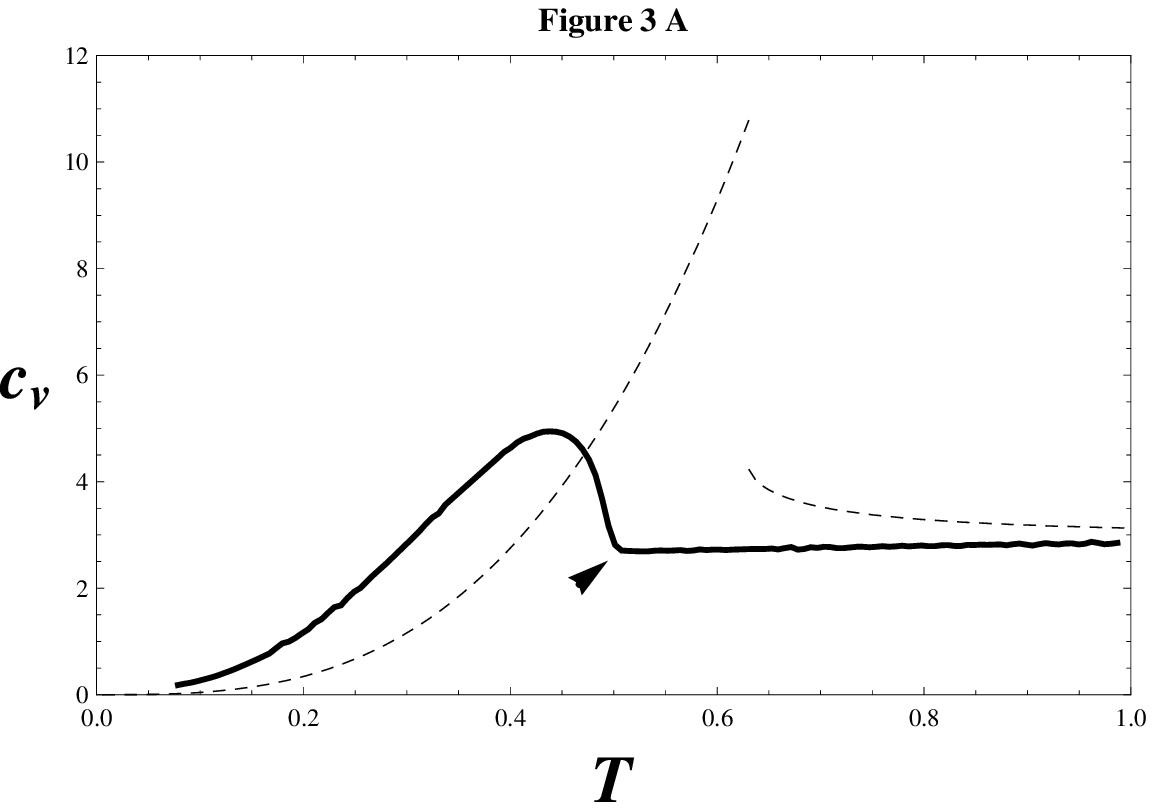}
  \end{center}
\end{figure}

\begin{figure}
  \begin{center}
  \includegraphics[scale=1]{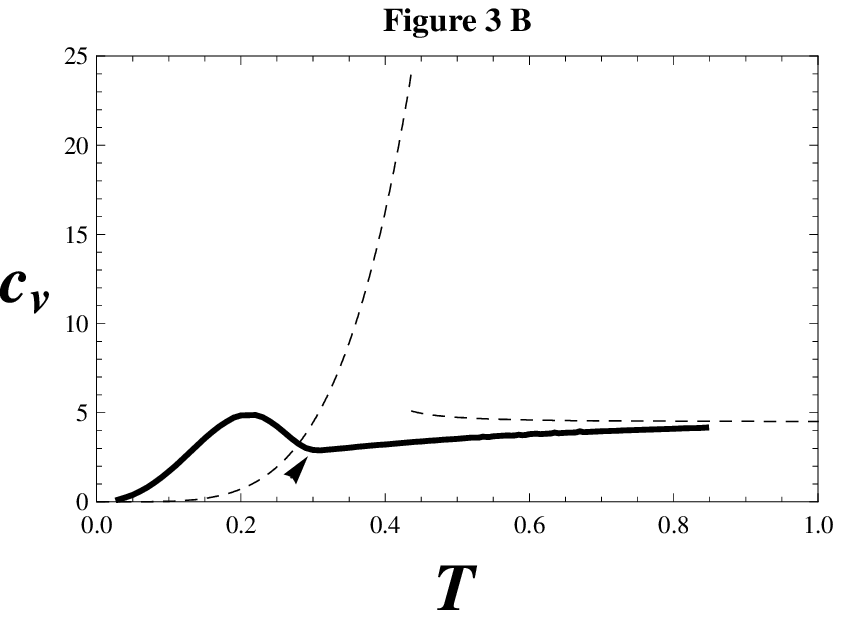}
  \end{center}
\end{figure}

\begin{figure}
  \begin{center}
  \includegraphics[scale=1]{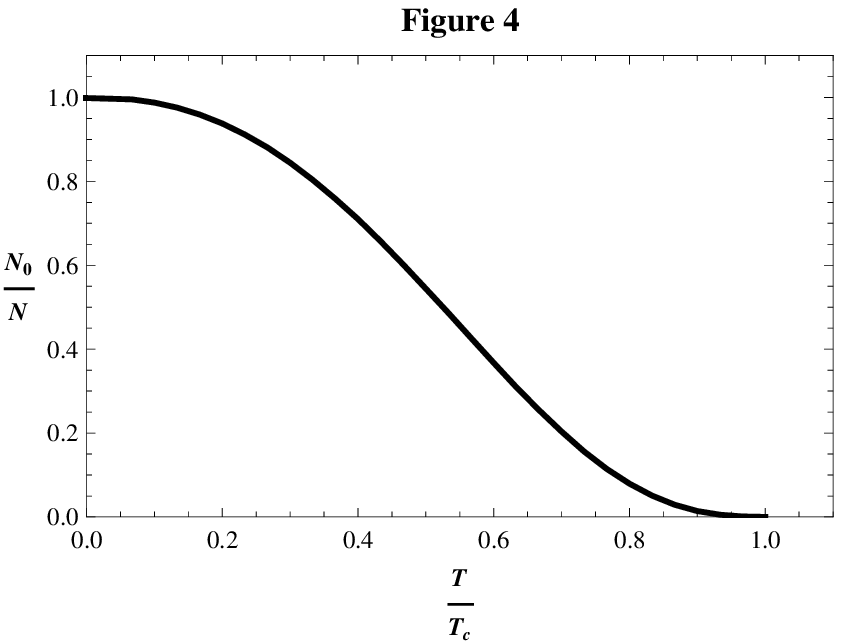}
  \end{center}
\end{figure}

Although no comprehensive experimental comparisons have been made using the present mechanical variables, Bagnato et al. have performed initial tests with a gas of $^{23}$Na atoms in a harmonic trap\cite{Magalhaes,Henn}.  Table I summarizes the comparison. In those experiments, the measured quantities are the parameters of the trap, the particle density, the temperature, the number of particles and the scattering length.  The experimental generalized pressure is calculated using Eq.(\ref{PV}) and the theoretical one is obtained with the HF approximation described above. Clearly, for temperatures above BEC the theory shows excellent agreement with experiments. For temperatures below $T_c$ - the last two entries of the Table- the Hartree-Fock approximation does not fare very well, as expected. It appears that the role of the interactions is still underestimated. An improvement using better theories, that include the proper role and statistical description of Bogoliubov modes\cite{Yukalov3} and/or including  the Popov approximation\cite{Dalfovo,Pethick}, is certainly desirable. Nevertheless, these initial experiments
 do show the usefulness of the
knowledge of the equation of state, not only for purposes of characterization, but clearly as an additional tool to learn about the elementary excitations of the superfluid state of the ultracold gases.

\begin{table}[htdp]
\begin{tabular}{|c|c|c|c|}
\hline \hline
 $\>\>\> T/\mu K \>\>\>$  &  $3 {\cal P V}/mN$ (exp) & $3{\cal PV}/mN$ (theo) & $\>\>\>
  T_c/\mu K \>\>\>$ \\

\hline\hline

 35 & 388 & 377.9 & 0.79  \\
 12 & 130 & 129.5 & 0.53 \\
 11.2 & 110 & 120.9 & 0.41 \\
 4.5 & 45.4 & 48.6 & 0.34 \\
 4.3 & 42.4 & 46.6 & 0.30 \\
 0.9 & 22.4 & 9.7 & 0.16 \\
 0.8 & 6.8 & 8.6 & 0.09 \\
 0.33 & 4.2 & 3.6 & 0.10 \\
 0.07 & 0.9 & 0.52 & 0.076 \\
 0.04 & 0.7 & 0.09 & 0.065 \\

\hline\hline
\end{tabular}\caption{Comparison of experimental generalized pressure ${\cal P}$ with a Hartree-Fock (HF) calculation for an ultracold Sodium gas, see Eq.(\ref{PV}).  ${\cal V} = 1/ \omega_x \omega_y \omega_z$ with $\omega_x = 2\pi \times 36.7$ Hz, $\omega_y = 2\pi \times 120.8$ Hz, and $\omega_z = 2\pi \times 159.6$ Hz\cite{Magalhaes}. The mass is $m = 23$ a. u. for Sodium and the scattering length is  $a = 65 a_0$ (Bohr radii). $N$ is the number of particles, which for all these experiments\cite{Henn} was from $N \sim 10^4$ to $N \sim 10^6$. The transition BEC temperatures have been calculated within the HF approach. }
\end{table}

\section{Final Remarks}

The main purpose of this brief review if to emphasize that the use of  the generalized thermodynamic variables here presented should lead to useful results for the analysis of the physics of ultracold {\it trapped} gases, by simply following the rules of thermodynamics. We have shown that, perhaps with an elaborate and lengthy experimental procedure, both the equation of state, ${\cal P} ={\cal P}({\cal V}/N,T)$, and the heat capacity, $C_{\cal V} = C_{\cal V}({\cal V}/N,T)$, can be readily measured with the current experimental setups. Knowledge of these two quantities suffices to know all the thermodynamics of the corresponding system. As we mention in Section II, extensions to other possible thermodynamic variables, such as the externally varied scattering length, may be readily incorporated. It is clear to us that for this framework to be really useful one needs, first, to change the usual intuition on volume and hydrostatic pressure to their generalized counterparts and, second, to provide examples where these variables lead to new insights. The latter is mainly a job for theory and in this review we have presented the study of a weakly interacting Bose gas within the Hartree-Fock approximation. The results are promising and compare well with experiments, but certainly improvement is needed specially for very low temperatures.

We have also shown that the present approach leads also to validate the use of LDA, although we have indicated that care must be taken when using it. This should be more delicate when dealing with phase-separated fluids where it is not clear if LDA suffices for their description since the interfacial widths of the phase boundaries are expected to be of the order of the range of the intermolecular
interactions\cite{RW}.  This comment may apply to the states found in Fermi trapped gases of $^6$Li atoms\cite{Ziewerlein,Hulet,Shin}, where there is evidence that the confined fluid phase-separates into a superfluid and a normal paramagnetic gas, showing an interfacial phase boundary. In general, for such inhomogeneous states, one should not expect a local picture to be valid across the interface; the thermodynamic
potentials are expected to
be non-local on the density profiles. An approach based on the generalized variables here analyzed may lead to a global and novel picture of those states.

\begin{acknowledgments}
Work supported by UNAM DGAPA IN-114308. We thank V.S. Bagnato and collaborators for stimulating discussions and for allowing us to use their experimental data.

\end{acknowledgments}

\appendix*

\section{Derivation of $I_2$}

In this Appendix we provide the derivation of second order contribution $I_2$ to the virial expansion of the grand potential, Eq.(\ref{Ome-vir}).

>From the general expressions in Section III, Eqs.(\ref{Omega})-(\ref{U2}), one finds,
\begin{eqnarray}
I_2 &=& \int d^3 r_1 \int d^3 r_2 \left( < \vec r_1, \vec r_2 | e^{-\beta H_2} |  \vec r_1, \vec r_2 > +
\epsilon < \vec r_1, \vec r_2 | e^{-\beta H_2} |  \vec r_2, \vec r_1 >  \right) - \nonumber \\
&&  \int d^3 r_1 < \vec r_1| e^{-\beta H_1} |  \vec r_1 > \>
\int d^3 r_2 <  \vec r_2 | e^{-\beta H_1} |   \vec r_2 >
\end{eqnarray}
where $\epsilon = \pm 1$ for bosons or fermions. The Hamiltonians $H_2$ and $H_1$ are given by Eqs.(\ref{HN}) and (\ref{H1}). We make the change of variables to center of mass and relative coordinates,
\begin{equation}
\vec R = (\vec r_1 + \vec r_2)/2 \>\>\>{\rm and} \>\>\>
\vec r = \vec r_1 - \vec r_2, \label{ch2}
\end{equation}
 with their canonical momenta $\vec P$ and $\vec p$, respectively. This gives,
 \begin{equation}
 H_2 = {\vec P^2 \over 2 (2m)} + V_{ext}(\vec R + {\vec r \over 2}) + V_{ext}(\vec R - {\vec r \over 2}) + {\cal H}_2 ,
 \end{equation}
 where
 \begin{equation}
{\cal H}_2 = {\vec p^2 \over 2 (m/2)} + u(r) \label{H2rel}
\end{equation}
is the two-particle relative coordinate Hamiltonian. The thermodynamic limit consists in approximating
\begin{equation}
V_{ext}(\vec R + {\vec r \over 2}) + V_{ext}(\vec R - {\vec r \over 2}) \approx 2 V_{ext}(\vec R) .\label{terlim}
\end{equation}
since the range of the center of mass motion becomes arbitrarily large as ${\cal V} \to \infty$, while the ensuing relative motion is bounded.  It is easy to verify that the latter is bounded by the interatomic interaction range $\sigma$ for large temperatures, as the motion is classical, while it is bounded by the thermal de Broglie wavelength or the scattering length, for low temperatures when the motion is quantum.

The above limit, Eq.(\ref{terlim}), separates the motion of the center of mass from that of the relative coordinates,
\begin{eqnarray}
I_2 &\approx& \left(\int d^3 R <\vec R | e^{-\beta({\vec P^2 \over 2 (2m)} + 2V_{ext}(\vec R))}|\vec R> \right) \times \nonumber \\
&& \int d^3 r \left(
< \vec r | e^{-\beta {\cal H}_2} |\vec r> +  \epsilon < \vec r | e^{-\beta {\cal H}_2 }|-\vec r> -
< \vec r | e^{-\beta p^2/2(m/2)}|\vec r> \right) . \label{I2aprox}
\end{eqnarray}
We note that the center of mass motion corresponds to a one-particle system of mass $nm$ moving in an external potential $nV_{ext}(\vec R)$. In the thermodynamic limit one is allowed to use the semiclassical density of states for the center of mass motion, yielding,
 \begin{equation}
I_2 =  {{\cal V} \over  \lambda_T^6} \zeta(2 \beta)  b_2(T)  \label{qb2},
 \end{equation}
where
\begin{equation}
{\cal V}\zeta(2\beta) = \int d^3 R \> e^{-2\beta V_{ext}(\vec R)} ,
\end{equation}
and where the quantum second virial coefficient is given by,
\begin{eqnarray}
b_2(T) &=& 2^{3/2} \lambda_T^3 \int d^3 r \left(
< \vec r | e^{-\beta {\cal H}_2} |\vec r> + \right. \nonumber \\
&&\left. \epsilon < \vec r | e^{-\beta {\cal H}_2 }|-\vec r> -
< \vec r | e^{-\beta p^2/2(m/2)}|\vec r> \right) . \label{qb2a}
\end{eqnarray}
As a rule, in the thermodynamic limit the center of mass motion is always quasiclassical\cite{LL}. The expression for the quantum second virial coefficient above can be seen to be the correct one by comparing, for instance, with the expression given in Ref.\cite{LL}. For slow collisions, the relevant ones for ultracold gases, $b_2$ depends on the scattering length $a$ and this may become quite large near a Feschbach or potential resonance. The formulae here derived may then not be applicable very near such a point, called the unitarity limit, but as it has been shown\cite{Ho} this may be expected since in such a limit  the system behaves as if near a critical point. We add that the description of the scattering al low energies near resonances is valid for interatomic potentials $u(r)$ that decay at least as $1/r^3$\cite{LLQM}.  In the classical limit one also finds that the interaction must be ``short-range",  otherwise $b_2$ does not exist.

To end this part, we find illustrative to calculate $b_2$ for an ideal quantum gas, i.e. for $u(r) = 0$. One finds the so-called ``exchange" contribution to the second virial coefficient:
\begin{equation}
b_2^{(0)} =  \epsilon \>  \frac{1}{2^{3/2}} \> \lambda_T^3.
 \label{b20}
\end{equation}

\newpage
\centerline{FIGURE CAPTIONS}

\medskip\noindent
{\bf Figure 1.} Phase diagram ${\cal P}-T$ of a weakly interacting Bose gas confined by a harmonic trap (A) and by a linear quadrupolar potential (B). Several isochores (${\cal V}/N =$ constant) are shown in dotted lines. The solid line shows the normal gas to superfluid (BEC) transition. Compare with Figure 2, the ideal case. In the latter the BEC transition line occurs at higher temperatures than in the interacting case. Note also that in the interacting case, below the transition temperature, the condensed phase exerts pressure.
Units are $\hbar = 1$, $m = 1$ and $a_0 = 1$.

\medskip\noindent
{\bf Figure 2.} Phase diagram ${\cal P}-T$ of an ideal Bose gas confined by a harmonic trap (A) and by a linear quadrupolar potential (B). Several isochores (${\cal V}/N =$ constant) are shown in dotted lines. The solid line shows the BEC transition. See caption of Figure 1 for further details.

\medskip\noindent
{\bf Figure 3} Specific heat at constant generalized (molar) volume $c_v$ vs. temperature $T$ for a gas confined by a a harmonic trap (A) and by a linear quadrupolar potential (B). The dotted line is the ideal case and the solid line the weakly interacting Bose gas. The transition temperature in the interacting case is marked with an arrow. Note that while in the ideal case the heat capacity is discontinuous at the transition temperature, it appears that in the interacting case it is completely continuous. Further, in the latter case, the transition does not occur at the maximum value of $c_v$ but at its {\it minimum}. Units are $\hbar = 1$, $m = 1$ and $a_0 = 1$.

\medskip\noindent
{\bf Figure 4}. Condensate fraction $N_0/N$ as a function of temperature $T$ for an interacting Bose gas confined by a linear quadrupolar trap. Note that the derivative of the curve is continuous at the transition. See text for details. Units are $\hbar = 1$, $m = 1$ and $a_0 = 1$.

\newpage
\centerline{\bf CORRESPONDING AUTHOR}

\medskip
\centerline{Dr. Victor Romero-Rochin}
\centerline{Instituto de Fisica, Universidad Nacional Autonoma de Mexico}
\centerline{Apartado Postal 20-364. 01000 Mexico, D.F. MEXICO}

\medskip
\centerline{Office: + 52 (55) 5622 - 5096}
\centerline{Fax: + 52 (55) 5622 - 2015}
\centerline{email: romero@fisica.unam.mx}

\end{document}